\documentclass[fleqn,usenatbib]{mnras}

\usepackage{newtxtext,newtxmath}

\usepackage[T1]{fontenc}
\usepackage{ae,aecompl}

\usepackage{graphicx}	
\usepackage{amsmath}	
\usepackage{amssymb}	
\usepackage{bm}		
\usepackage{subfig}	

\newcommand\plk{\textit{Planck~}}

\title[Introducing constrained matched filters]{Introducing constrained matched filters for improved separation of point sources from galaxy clusters}

\author[Erler, Ramos--Ceja, Basu \& Bertoldi]{Jens Erler$^{1}$\thanks{E-mail: jens@astro.uni-bonn.de}
\thanks{Member of the International Max Planck Research School (IMPRS) for Astronomy and Astrophysics at the Universities of Bonn and Cologne.},
{Miriam E. Ramos--Ceja$^{1}$,}
{Kaustuv Basu$^{1}$ and}
{Frank Bertoldi$^{1}$}\\
$^{1}$Argelander-Institut f{\"u}r Astronomie, Universit{\"a}t Bonn, Auf dem H{\"u}gel 71, 53121 Bonn, Germany}

\date{Accepted XXX. Received YYY; in original form ZZZ}

\pubyear{2019}

\begin{document}
\label{firstpage}
\pagerange{\pageref{firstpage}--\pageref{lastpage}}
\maketitle

\begin{abstract}
Matched filters (MFs) are elegant and widely used tools to detect and measure signals that resemble a known template in noisy data. However, they can perform poorly in the presence of contaminating sources of similar or smaller spatial scale than the desired signal, especially if signal and contaminants are spatially correlated. We introduce new multicomponent MF and matched multifilter (MMF) techniques that allow for optimal reduction of the contamination introduced by sources that can be approximated by templates. The application of these new filters is demonstrated by applying them to microwave and X-ray mock data of galaxy clusters with the aim of reducing contamination by point-like sources, which are well approximated by the instrument beam. Using microwave mock data, we show that our method allows for unbiased photometry of clusters with a central point source but requires sufficient spatial resolution to reach a competitive noise level after filtering. A comparison of various MF and MMF techniques is given by applying them to \textit{Planck} multifrequency data of the Perseus galaxy cluster, whose brightest cluster galaxy hosts a powerful radio source known as Perseus A. We also give a brief outline how the constrained MF (CMF) introduced in this work can be used to reduce the number of point sources misidentified as clusters in X-ray surveys like the upcoming eROSITA all-sky survey. A {\small PYTHON} implementation of the filters is provided by the authors of this manuscript at \url{https://github.com/j-erler/pymf}.
\end{abstract}




\section{Introduction}

Matched filtering (MF) is a technique for the extraction of the flux of sources with a well-known spatial template at optimal signal-to-noise ratio (SNR).  MF was first proposed for the study of the kinetic Sunyaev--Zeldovich (kSZ) signal from clusters of galaxies by \citet{Haehnelt96} and subsequently developed and generalized by \citet{Herranz02} and \citet{Melin06} for the extraction of the thermal Sunyaev--Zeldovich (tSZ) signal from multifrequency data sets like those delivered by the \plk mission, giving rise to what is now known as the matched multifilter (MMF). These filters have since been adopted with great success by the SPT, ACT and Planck Collaborations to extract the tSZ signal of clusters from their respective multifrequency data sets (\citealt{Hasselfield13, Bleem15, Planck16}).

While MFs perform admirably in separating diffuse Galactic foregrounds and primary cosmic microwave background (CMB) anisotropies from the SZ signal of clusters, contamination by point sources remains an issue (e.g. \citealt{Bartlett06, Melin06}) and can lead to significant biases in the measured cluster parameters \citep{Knox04, Aghanim05, Lin07, Sehgal10}. This problem is mitigated to a degree by MMFs due to the prior knowledge of the tSZ spectrum that is used to construct these multifilters, but accurate photometry of clusters that contain a central radio source remains challenging. Point source confusion is also a central concern for the detection of clusters in X-ray observations (e.g. \citealt{Biffi18, Koulouridis18}).
\citet{Tarrio16} and {Tarrio18} recently demonstrated that point source confusion can be reduced by a joint SZ and X-ray MMF analysis, using their very different spectral characteristics at microwave frequencies compared to the X-ray regime. 

In this work, we present a  multicomponent extension of the MF concept that can improve the separation of contaminants that can be approximated by well-known templates (e.g. point sources) based purely on their spatial characteristics. This approach is mathematically identical to the generalized multicomponent internal linear combination (ILC) algorithms introduced by \citet{Remazeilles11a, Remazeilles11b} and \citet{Hurier13}, which can be thought of as MFs in frequency space, and allows for an unbiased photometry of clusters with a central point source. Generalizing our method to multifrequency data gives rise to a new matched multifiltering technique that combines spatial and spectral constraints to provide an optimal separation. A similar but less general approach was presented by \citet{Herranz05}, who showed that the tSZ and kSZ signals of clusters can be separated with MMFs that use the different spectra of the two effects but take the same spatial template for the two components, which restricts the method from being applied to other contaminating sources. In this work, we derive our new filters and demonstrate their application using mock microwave and X-ray data of clusters, as well as \textit{Planck} data of the Perseus galaxy cluster.

This article is structured as follows: Section~\ref{sec:matched_filtering} introduces MFs and MMFs for galaxy clusters and our proposed constrained filters in detail. Section~\ref{sec:simulations} describes our simulation pipeline for the creation of mock data that are used to test the performance of the constrained MFs. The results obtained on both simulations and on data from the \textit{Planck} mission are presented in Section~\ref{sec:results}. In Section~\ref{sec:discussion}, we provide a discussion of our new technique and give an outlook to its application in future experiments. Section~\ref{sec:conclusion} provides a summary and concludes our analysis. 

Throughout this paper we assume a flat Lambda cold dark matter cosmology with $\Omega_\Lambda = 0.7$, $\Omega_\mathrm{b} = 0.05$, $h = 0.7$, and $T_\mathrm{CMB} = 2.7255 \, \mathrm{K}$.
$E(z) \equiv H(z)/H_0 = (\Omega_\mathrm{m}(1+z)^3 + \Omega_\Lambda)^{1/2}$ denotes the redshift-dependent Hubble ratio and $\rho_\mathrm{crit}(z) = 3H(z)^2/(8 \pi G)$ the critical density of the Universe at redshift $z$.
Unless noted otherwise, the quoted parameter uncertainties refer to the 68 per cent credible interval. All-sky maps were processed with {\small HEALPIX} (v3.31; \citealt{Goerski05}).

\section{Matched filtering}
\label{sec:matched_filtering}

Setting up an MF requires only very limited knowledge about the astrophysical content of a data set. We assume that an observed map $\mathbfss{I}_{\nu}$ at frequency $\nu$ represents a linear combination of the desired signal, e.g. the SZ signal from galaxy clusters with the spectrum $f(\nu)$, plus a noise map $\mathbfss{N}_{\nu}$ that contains both instrumental noise and astrophysical emission:  
\begin{equation}
	\mathbfss{I}_{\nu} = f(\nu) \cdot A \, \mathbfss{y} + \mathbfss{N}_{\nu}.
    \label{eq:skymodel}
\end{equation}
The signal must be well approximated by a known spatial template $\mathbfss{y}$ like the projected pressure profile of clusters. We now would like to construct a filter $\bm{\Psi}$ that returns the signal (i.e. the amplitude $A$ of the source template if $\mathbfss{y}$ is normalized to unity) at maximum significance.
Using the flat sky approximation and changing to Fourier space, an MF $\bm{\Psi}$ can be constructed by minimizing the variance of the filtered map (e.g. \citealt{Schaefer06})
  \begin{equation}
   \sigma^2 = \bm{\Psi}^\mathrm{T} \mathbfss{C} \bm{\Psi},
   \label{eq:variance}
  \end{equation}
where $\mathbfss{C}$ is the azimuthally averaged noise power spectrum of the unfiltered map expressed as a diagonal matrix \hbox{$\mathbfss{C} = \mathrm{diag}(|\mathbfss{N}(\bm{k})|^2)$}. Here $\bm{k}$ denotes the two-dimensional spatial frequency that corresponds to the two-dimensional sky position $\bm{x}$ in Fourier space. At the same time, we demand the filtered field to be an unbiased estimator of the deconvolved amplitude of the signal template at the position of sources. This condition can be written as
  \begin{equation}
   \bm{\Psi}^{T}\bm{\tau} = 1,
  \end{equation}
where $\bm{\tau}$ is the Fourier transform of the source template $\mathbfss{y}$ convolved with the instrument beam. A solution to this optimization problem is found by introducing a Lagrange multiplier $\lambda$, which leads to a system of linear equations
\begin{equation}
   \begin{pmatrix}
    2\cdot \mathbfss{C}  & -\bm{\tau} \\
    \bm{\tau}^\mathrm{T} & 0
   \end{pmatrix} 
   \begin{pmatrix}
    \bm{\Psi} \\
    \lambda 
   \end{pmatrix}
   =
   \begin{pmatrix}
    0 \\
    1 
   \end{pmatrix},
   \label{eq:lin_eq}
 \end{equation}
the solution to which is:
\begin{equation}
  \bm{\Psi} = \left[ \bm{\tau}^\mathrm{T} \mathbfss{C}^{-1}\bm{\tau} \right]^{-1} \bm{\tau}\mathbfss{C}^{-1}.
  \label{eq:mf}
\end{equation}
The MF derived here is optimal in the least square sense and was first proposed for the study of galaxy clusters by \citet{Haehnelt96}.
Although it is most commonly applied to data sets with Gaussian noise, Gaussianity is not a strict requirement. Non-Gaussian noise will not cause a bias but the solution might no longer be optimal \citep{Melin06}. However, optimal MFs were recently derived for the low-number count Poisson noise regime that is relevant for X-ray and $\gamma$-ray observations \citep{Ofek18, Vio18}.

\subsection{Constrained matched filters (CMF)}
\label{sec:constrained_matched_filtering}

\begin{figure*}
  \centering
  \subfloat{%
  \includegraphics[width=0.48\textwidth]{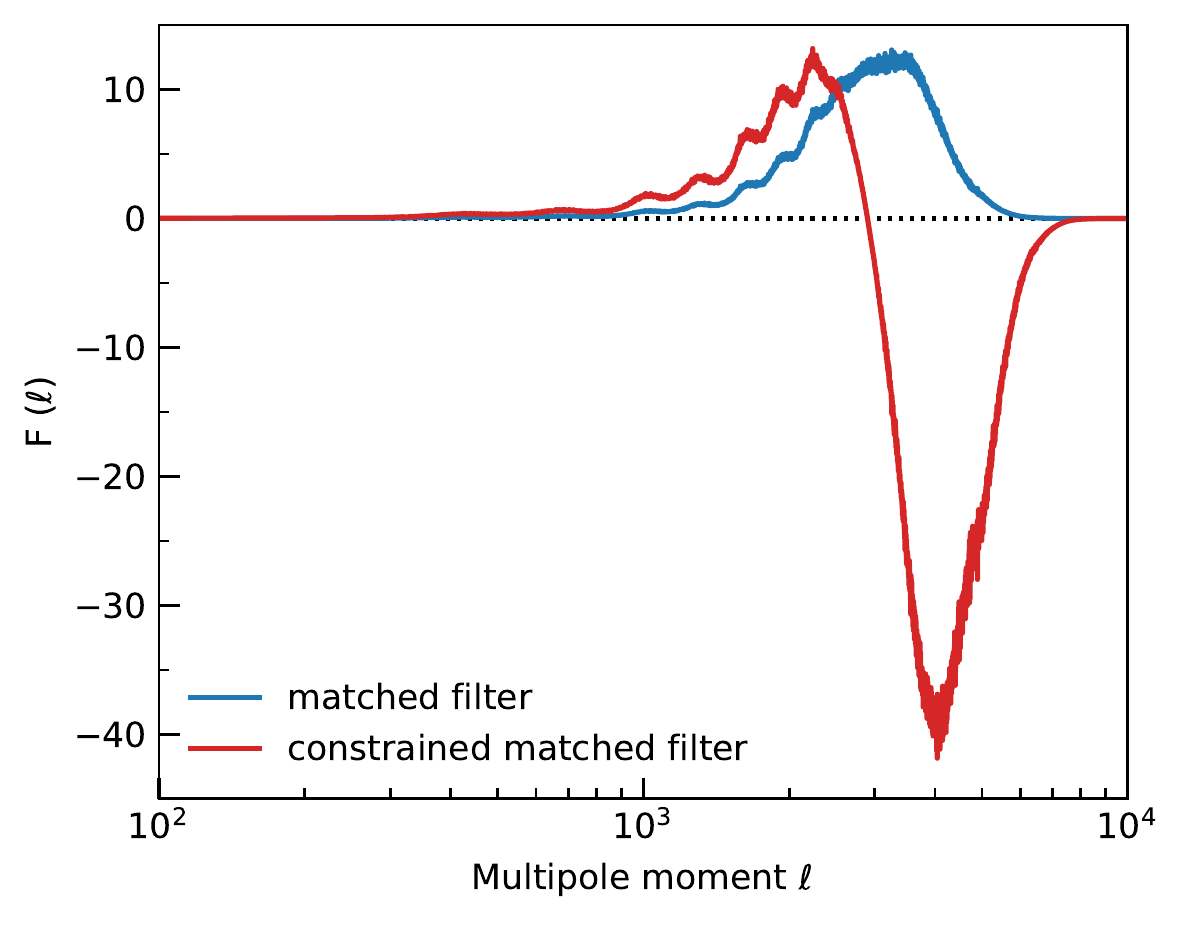}}
  \qquad
  \subfloat{%
  \includegraphics[width=0.48\textwidth]{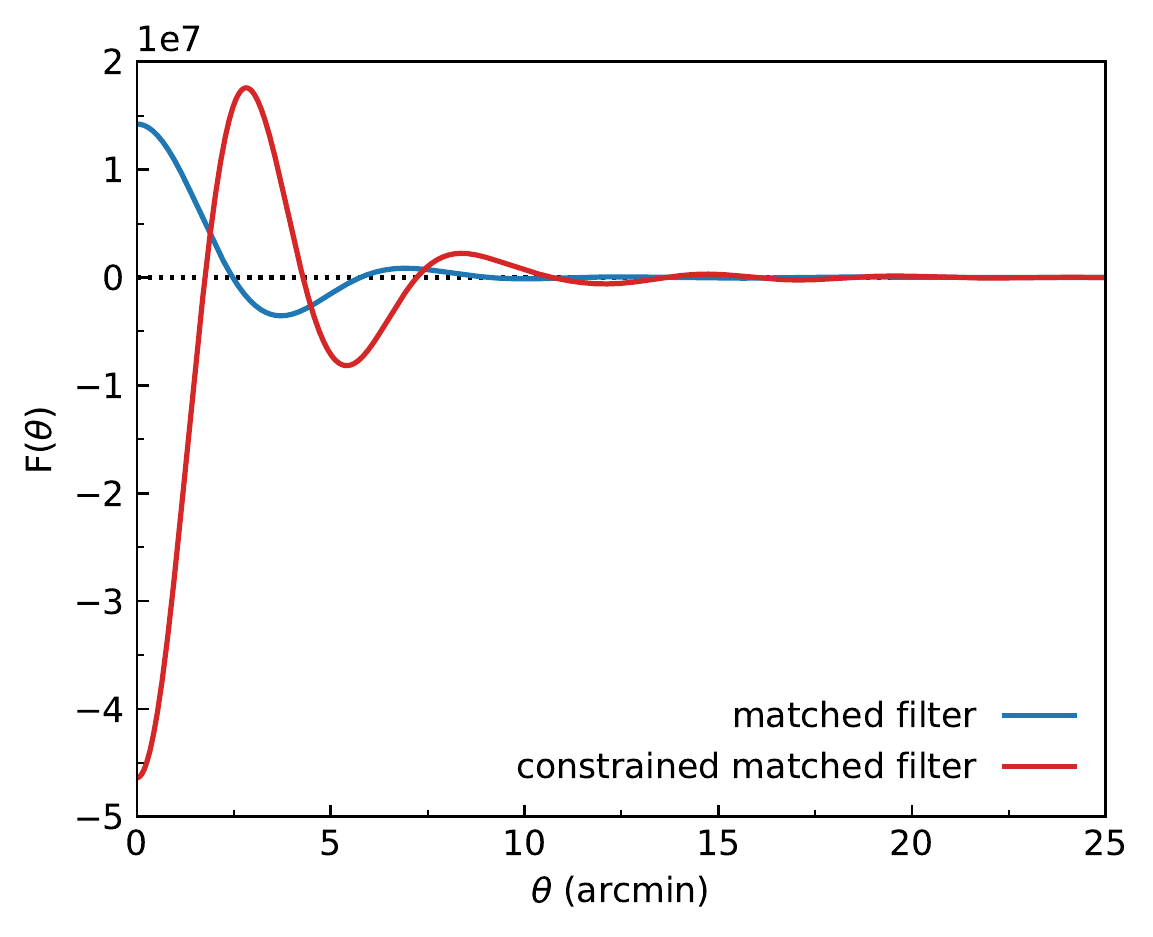}}
  \caption{Comparison of the traditional MF and the CMF presented in this work. The filters are computed using the all-sky formalism given in appendix~\ref{sec:allsky} on the simulated $150 \, \mathrm{GHz}$ sky presented in Section~\ref{sec:simulations} assuming a $\beta$-model \citep{Cavaliere76} with core radius $\theta_c = 2 \, \mathrm{arcmin}$ and $\beta = 1$ convolved with a Gaussian beam with an FWHM of $5 \, \mathrm{arcmin}$ as the source template. A point source template has been used as an additional constraint for the CMF. \textbf{Left-hand panel:} Filter window functions in spherical harmonic space. \textbf{Right-hand panel:} Filter kernel profiles in real space.}
  \label{fig:cmf_example}
\end{figure*}

We now show that the MF concept can be generalized to multiple sources with known spatial templates. For this we assume that the observed sky is a linear combination of $n$ sources with known templates $\mathbfss{y}_i$ plus noise:
\begin{equation}
	\mathbfss{I}_{\nu} = f_1(\nu) \cdot A_1 \, \mathbfss{y}_{1} + \dots + f_n(\nu) \cdot A_n \, \mathbfss{y}_n + \mathbfss{N}.
\end{equation}
Our goal is to construct a filter that minimizes the variance of the filtered map as defined in equation~(\ref{eq:variance}) and at the same time has an unbiased response to the chosen source template. We now place additional constraints by e.g. demanding the filter to have zero response to contaminating sources with well-known spatial templates:
\begin{align}
  \begin{split}
      \bm{\Psi}^{T}\bm{\tau}_1 &= 1 \\
      \bm{\Psi}^{T}\bm{\tau}_2 &= 0 \\
      &\vdots \\
      \bm{\Psi}^{T}\bm{\tau}_n &= 0.
  \end{split}
  \label{eq:constraints_cmf}
\end{align}
In the following it is convenient to construct a matrix $\mathbfss{T}$ of dimensions $n_k \times n$ from the $n$ spatial templates $\bm{\tau}_i$:
\begin{equation}
  \mathbfss{T}=\begin{pmatrix}
    \tau_1[1]  & \tau_2[1] & \dots & \tau_n[1]  \\
    \vdots      & \vdots       & \ddots& \vdots \\
    \tau_1[n_k] & \tau_2[n_k] & \dots & \tau_n[n_k] \\
    \end{pmatrix}.
\end{equation}
We can derive the form of the new filter by solving a system of linear equations analogous to equation~(\ref{eq:lin_eq})
\begin{equation}
   \begin{pmatrix}
    2\cdot \mathbfss{C}  & -\mathbfss{T} \\
    \mathbfss{T}^\mathrm{T} & 0
   \end{pmatrix} 
   \begin{pmatrix}
    \bm{\Psi} \\
    \bm{\lambda} 
   \end{pmatrix}
   =
   \begin{pmatrix}
    0 \\
    \bm{e} 
   \end{pmatrix},
 \end{equation}
where $\bm{e} = (1,0,\dots)^\mathrm{T}$ is a vector that contains the response of the filter to the $n$ constraints defined in equation~(\ref{eq:constraints_cmf}) and $\bm{\lambda}$ are the $n$ Lagrange multipliers.
The solution for the CMF is
\begin{equation}
  \bm{\Psi} = \bm{e}^\mathrm{T}\left[\mathbfss{T}^\mathrm{T} \mathbfss{C}^{-1}\mathbfss{T} \right]^{-1} \mathbfss{T} \mathbfss{C}^{-1},
\end{equation}
which is similar to the one of the traditional MF in equation~(\ref{eq:mf}). A possible application of this new filter is the reduction of point source contamination in observations of galaxy clusters, which will be explored in Section~\ref{sec:results}. However any other contaminating source with a well-known template or even multiple sources could be set to zero using this approach. This benefit will come at the cost of a reduced SNR, which will be discussed in Section~\ref{sec:results}. A comparison of the two filters using simulated microwave data of galaxy clusters and point sources is shown in Fig.~\ref{fig:cmf_example}.

A mathematically identical multicomponent generalization to the CMF has been derived and successfully applied for ILC algorithms \citep{Remazeilles11a, Remazeilles11b, Hurier13}, which are commonly used to extract Comptonization maps from \textit{Planck} data using the spectrum of the tSZ signal while zeroing out the primary CMB anisotropies by constraining their well understood blackbody spectrum. 
\subsection{Constrained matched multifilters (CMMF)}

Both the MF and CMF presented previously were built to be applied to a single-frequency map. However, the MF concept can be generalized to multifrequency data sets like the ones delivered by \textit{Planck}. These generalized techniques are known as matched multifilters (MMF; \citealt{Herranz02, Melin06, Lanz10, Melin12, Tarrio16, Tarrio18}) and are designed to use prior spatial and spectral information about a source to return an optimally filtered map of A in the least-square sense. We will show here that the MMF concept can be generalized to separate multiple components with known spatial and spectral templates in an analogous way to the single-frequency filter. We start again by constructing a simple model of the observed sky. As before, we can represent observations of the sky as a linear mixture of astrophysical emission and noise:
\begin{equation}
	\bm{I}(\bm{x}) = \bm{f}_\nu \cdot A \, y(\bm{x}) + \bm{N}(\bm{x}).
\end{equation}
Different from equation~(\ref{eq:skymodel}) we now describe the observed maps as vectors in frequency space with $n_\nu$ components at each sky position $\bm{x}$ in order to simplify the notation. Using this formalism and changing to Fourier space, the multifrequency source template will be given at each $\bm{k}$ as a vector $\bm{F}$ in frequency space
\begin{equation}
	\bm{F}(\bm{k}) = \bm{f}_{\nu} \, y(\bm{k}) \, B_\nu(\bm{k}),
\end{equation}
where $B_\nu(\bm{k})$ denotes the Fourier transform of the beam, which in general will be frequency dependent.
We now aim to find a filter $\bm{\Psi}(\bm{k})$ that, as before, has unit response to the multifrequency source template:
\begin{equation}
	\int \mathrm{d}^2k \ \bm{\Psi}^{T}(\bm{k})\bm{F}(\bm{k}) = 1.
\end{equation}
We therefore construct a series of $n_\nu$ filters $\Psi(\bm{k})$ that are the components of $\bm{\Psi}(\bm{k})$. The final result will be a single map that is the linear combination of the observed maps, each convolved with their respective frequency-dependent filter. The MMF is derived analogously to the single-frequency case by demanding minimum variance of the filtered map at each spatial scale
  \begin{equation}
   \sigma^2 = \bm{\Psi}^\mathrm{T}(\bm{k}) \mathbfss{P}(\bm{k}) \bm{\Psi}(\bm{k}),
   \label{eq:variance_MMF}
  \end{equation}
where $\mathbfss{P}$ is the noise power spectrum, a matrix in frequency space with $n_\nu \times n_\nu$ components for each $\bm{k}$ that are defined as $P_{ij}(\bm{k})\delta(\bm{k}-\bm{k}') = \langle N_{\nu_{i}}(\bm{k})N_{\nu_{j}}^{*}(\bm{k})\rangle$. The asterisk denotes the complex conjugate. The MMF is then given by
\begin{equation}
  \bm{\Psi}(\bm{k}) = \sigma_\mathrm{MMF}^2 \mathbfss{P}^{-1}(\bm{k}) \bm{F}(\bm{k}),
\end{equation}
with the variance of the filtered map:
\begin{equation}
  \sigma_\mathrm{MMF}^2 = \left[ \int \mathrm{d}^2k \ \bm{F}^\mathrm{T}(\bm{\bm{k}}) \mathbfss{P}^{-1}(\bm{k}) \bm{F}(\bm{k}) \right]^{-1}.
\end{equation}
The MMF derived here was employed with great success for the detection and photometry of galaxy clusters by the ACT, SPT, and Planck collaborations (\citealt{Hasselfield13, Bleem15, Planck16}). 

We now show that a CMMF can be constructed in similar fashion as before. The aim is to find a filter that allows us to constrain multiple well-known multifrequency source templates to reduce the impact of well-characterized contaminants on the filtered map. We begin by assuming that the observed sky is a linear mixture of known sources plus noise:
\begin{equation}
	\bm{I}(\bm{x}) = \bm{f}_{\nu,1} \cdot A_1 \, y_1(\bm{x}) + \dots + \bm{f}_{\nu,n} \cdot A_n \, y_n(\bm{x}) + \bm{N}(\bm{x}).
\end{equation}
Next, we define the desired response of the filter to our known source templates:
\begin{align}
  \begin{split}
      \int \mathrm{d}^2k \ \bm{\Psi}^{T}(\bm{k})\bm{F}_1(\bm{k}) &= 1 \\
      \int \mathrm{d}^2k \ \bm{\Psi}^{T}(\bm{k})\bm{F}_2(\bm{k}) &= 0 \\
      &\vdots \\
      \int \mathrm{d}^2k \ \bm{\Psi}^{T}(\bm{k})\bm{F}_n(\bm{k}) &= 0.
  \end{split}
  \label{eq:constraints}
\end{align}
For each $\bm{k}$ the constraints can be written as a matrix $\mathbfss{U}$ with dimensions $n_\nu \times n$:
\begin{equation}
  \mathbfss{U}(\bm{k})=\begin{pmatrix}
    F_1[1](\bm{k})  & F_2[1](\bm{k}) & \dots & F_n[1](\bm{k})  \\
    \vdots      & \vdots       & \ddots& \vdots \\
    F_1[n_\nu](\bm{k}) & F_2[n_\nu](\bm{k}) & \dots & F_n[n_\nu](\bm{k}) \\
    \end{pmatrix}.
\end{equation}
By minimizing the variance of the filtered map, we find that the CMMF is
\begin{equation}
  \bm{\Psi}(\bm{k}) = \bm{e}^\mathrm{T} \mathbfss{S}^{-1} \mathbfss{P}^{-1}(\bm{k}) \mathbfss{U}(\bm{k}),
\end{equation}
with the $n\times n$ matrix $\mathbfss{S}$ defined as:
\begin{equation}
  \mathbfss{S} = \int \mathrm{d}^2k \ \mathbfss{U}^\mathrm{T}(\bm{k}) \mathbfss{P}^{-1}(\bm{k}) \mathbfss{U}(\bm{k}).
\end{equation}
The variance of the filtered map can be computed as:
\begin{equation}
  \sigma_\mathrm{CMMF}^2 = \int \mathrm{d}^2k \bm{\Psi}^\mathrm{T}(\bm{k}) \mathbfss{P}(\bm{k}) \bm{\Psi}(\bm{k}) .
\end{equation}

The CMMF defined this way can be used to separate sources in a similar fashion as the single frequency CMF, but for multifrequency data sets. This will require both a spatial and a spectral template for each constrained source, e.g. if we want to extract galaxy clusters from multifrequency microwave data while minimizing point source contamination we need to know the beam as well as the spectral energy distribution (SED) of the point sources. This makes the method very efficient in cleaning the data, but it will be limited to a specific type of source. Reducing the contamination of radio and far-infrared point sources at the same time can be achived by placing two additional constraints using the same spatial template (i.e. the beam) but two different SEDs. We will compare the performance of the different filters presented here by applying them to \textit{Planck} High Frequency Instrument (HFI) data of the Perseus galaxy cluster in Section~\ref{sec:results}. 

\section{Simulations}
\label{sec:simulations}
\subsection{The SZ effect of galaxy clusters}

In order to test the performance of the CMF and compare it to the traditional MF we prepared a pipeline for the creation of mock images of the microwave sky. We use the tSZ effect signal \citep{Sunyaev70, Sunyaev72, Birkinshaw99, Carlstrom02} of galaxy clusters as our sources of interest.

The tSZ effect is a secondary anisotropy of the CMB that is caused by inverse Compton scattering of CMB photons by free electrons in the intracluster medium (ICM). The tSZ effect causes a characteristic distortion of the CMB spectrum with a temperature decrement at low $(\lesssim 217 \, \mathrm{GHz})$ and a temperature increment at high $(\gtrsim 217 \, \mathrm{GHz})$ frequencies. Peculiar motion of clusters will cause a red/blue-shift of the CMB in their rest frame, which gives rise to the kSZ effect. The spectra of the SZ signals are commonly expressed as a temperature shift relative to the CMB monopole, which can be written as
\begin{equation}
 \frac{\Delta T_\mathrm{SZ}}{T_\mathrm{CMB}} = \underbrace{f(x, T_\mathrm{e}) \, y}_{\mathrm{tSZ}} - \underbrace{\tau_\mathrm{e} \, \left(\frac{\varv_\mathrm{pec}}{c}\right)}_\mathrm{kSZ},
\end{equation}
where $T_\mathrm{CMB}$ is the CMB temperature, $c$ is the speed of light, $\varv_\mathrm{pec}$ is the peculiar velocity along the line of sight, $f(x, T_\mathrm{e})$ is the relativistic tSZ (rSZ) spectrum \citep[e.g.,][]{Wright79, Itoh98, Chluba12}, \hbox{$x \equiv h \nu/(k_\mathrm{B}T_\mathrm{CMB})$} is the dimensionless frequency, \hbox{$\tau_\mathrm{e}(r) = \sigma_\mathrm{T} \int n_\mathrm{e}(r) \, \mathrm{d}l$} is the optical depth of the plasma and $y$ is the Comptonization parameter:
\begin{equation}
y(r) = \frac{\sigma_\mathrm{T}}{m_\mathrm{e}c^2}\int_\mathrm{l.o.s.} \mathrm{d}l \, \underbrace{n_\mathrm{e}(r) k_\mathrm{B} T_\mathrm{e}(r)}_{P_\mathrm{e}(r)}.
\end{equation}
Here, $k_\mathrm{B}$ is the Boltzmann constant, $\sigma_\mathrm{T}$ is the Thomson cross-section, $m_\mathrm{e}$ is the electron rest mass and $n_\mathrm{e}$ and $T_\mathrm{e}$ are the number density and temperature of the electrons in the ICM. 

The Comptonization parameter is a measure of the gas pressure integrated along the line of sight (l.o.s.) and is computed by projection of the Generalized Navarro--Frenk--White (GNFW) pressure profile \citep{Nagai07} using the parametrization presented by \citet{Arnaud10}
\begin{equation}
 \frac{P_\mathrm{e}(r)}{\mathrm{keV \, cm^{-3}}} = 1.65 \times 10^{-3} E(z)^{8/3} \left(\frac{M_{500}}{3\times10^{14} \, \mathrm{\mathrm{M_\odot}}}\right)^{0.79} p\left( \frac{r}{r_{500}} \right),
\end{equation}
where $p(r/r_{500})$ is the so-called `universal' shape of the cluster pressure profile
\begin{equation}
 p(r) = \frac{P_0}{\left(c_{500} \frac{r}{r_{500}} \right)^\gamma \left[1+\left(c_{500} \frac{r}{r_{500}} \right)^\alpha \right]^{(\beta-\gamma)/\alpha}},
\end{equation}
for which we adopt the best-fit values for the profile parameters \hbox{$P_0,c_{500},\gamma,\alpha, \ \mathrm{and} \ \beta$} presented by \citet{Arnaud10}. In the following we refer to this profile as the GNFW profile. The characteristic cluster size $r_{500}$ marks the radius of the sphere within which the average matter density is 500 times the critical density, while $M_{500}$ is the total mass enclosed within $r_{500}$.
The temperature profile of the clusters is computed assuming a polytropic relation $n_\mathrm{e} T_\mathrm{e} = n_\mathrm{e}^\delta$ between electron density and temperature, with $\delta = 1.2$ \citep{Ostriker05}.

\subsection{Simulating the microwave sky}

The simulated clusters are added to an artificial CMB map computed from a synthetic power spectrum that was generated using {\small CAMB} \citep{Lewis00}. We account for emission from the cosmic infrared background (CIB) by adding maps of the resolved and the clustered CIB 

\begin{figure*}
  \centering
  \includegraphics[width=0.8\textwidth]{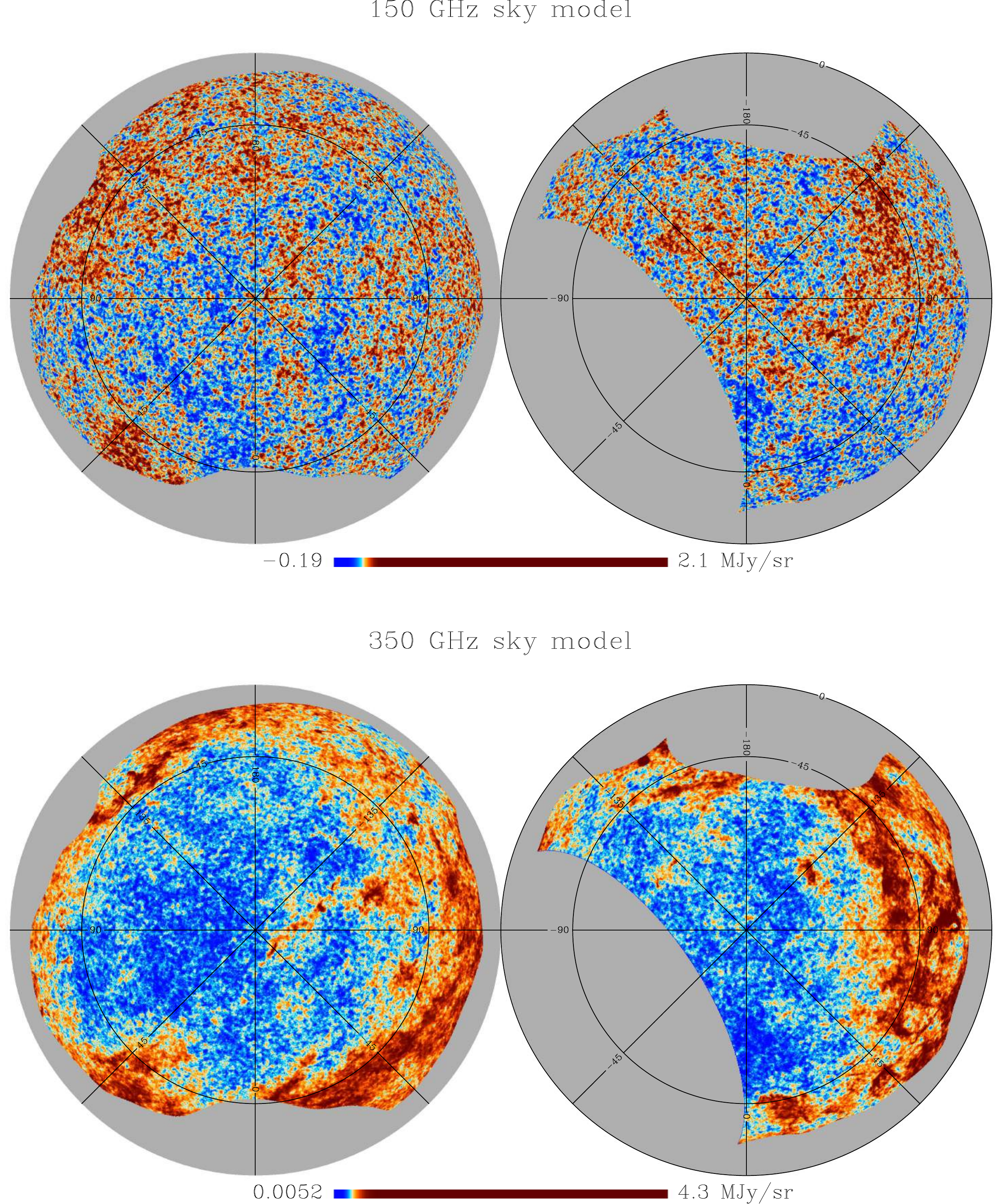}
\caption{Orthographic view of the simulated microwave sky at $150$ and $350 \, \mathrm{GHz}$ that was used to test the different filtering techniques. The projection is centred on the Galactic north and south pole. The maps are shown in histogram equalized scale to enhance the dynamic range. The composition of the maps is described in Section~\ref{sec:simulations}. We remove the brightest parts of the Galactic disc by applying a 40 per cent Galactic mask and exclude the part of the sky that has not been observed by the NVSS.}
  \label{fig:sky_model}
\end{figure*}
\noindent provided by the WebSky Extragalactic CMB Mocks team\footnote{The mocks are provided at \url{https://mocks.cita.utoronto.ca}} to our simulation pipeline.
We use the {\small PYTHON} sky model ({\small PSM}; \citealt{Thorne17}) to obtain maps of Galactic synchrotron, free--free, spinning dust, and thermal dust components. The PSM uses the most recent foreground maps published by the \citet{Planck_foregrounds} for the latter three components and adds small-scale fluctuations to all maps following an approach similar to the one  presented by \citet{Miville07}. 

Compact radio sources are modelled by including all sources from the NVSS point source catalogue \citep{Condon98}. The measured fluxes densities at 1.4 GHz are extrapolated to microwave frequencies assuming a power-law SED, $I(\nu) \propto \nu^{-\alpha}$ with a spectral index $\alpha$ randomly drawn for each source from a Gaussian distribution with a mean of 0.5 and a standard deviation of 0.1. Galactic and extragalactic near-infrared point sources are included by adding the sources listed in the IRAS point source catalogue \citep{Beichman88} by following the approach presented by \citet{Delabrouille13} to extrapolate the reported flux densities to lower frequencies.  

\begin{figure*}
  \centering
  \subfloat{%
  \includegraphics[width=0.48\textwidth]{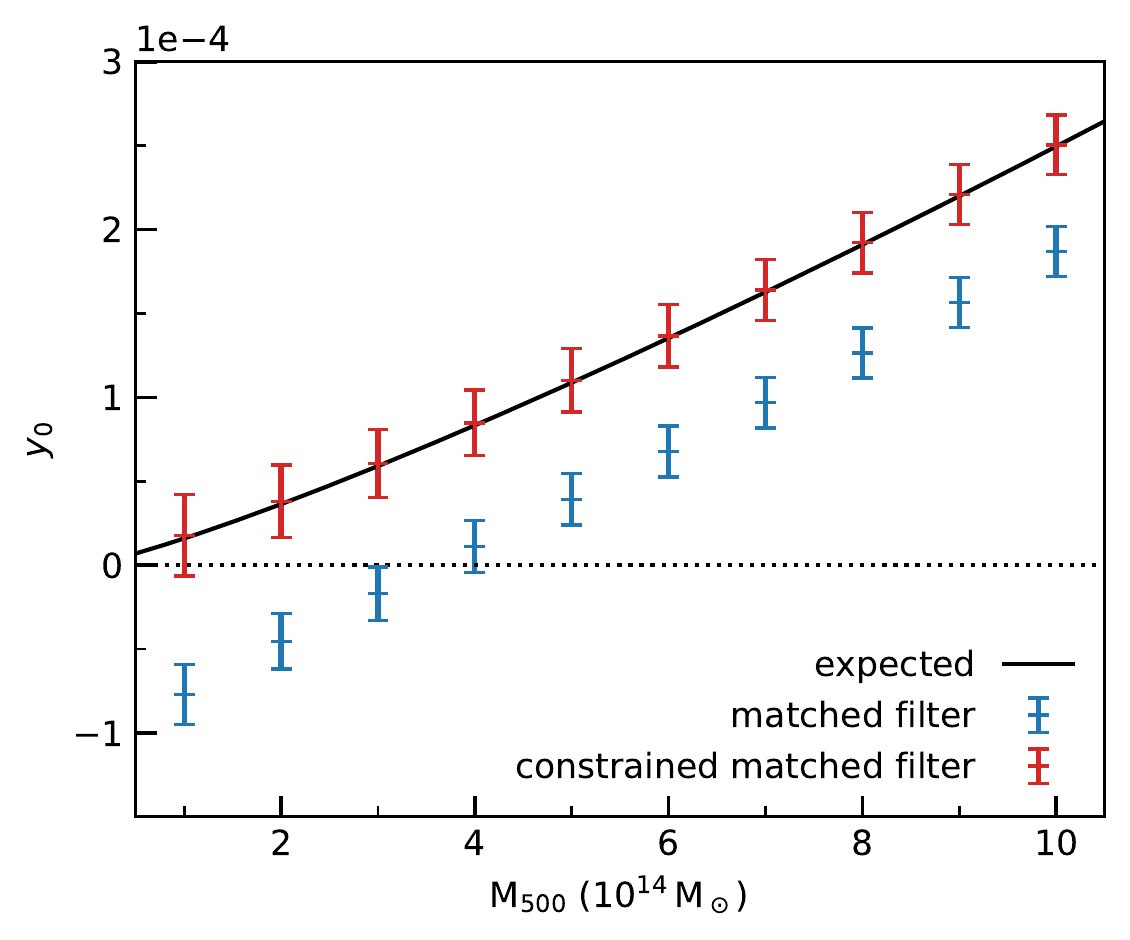}}
  \qquad
  \subfloat{%
  \includegraphics[width=0.48\textwidth]{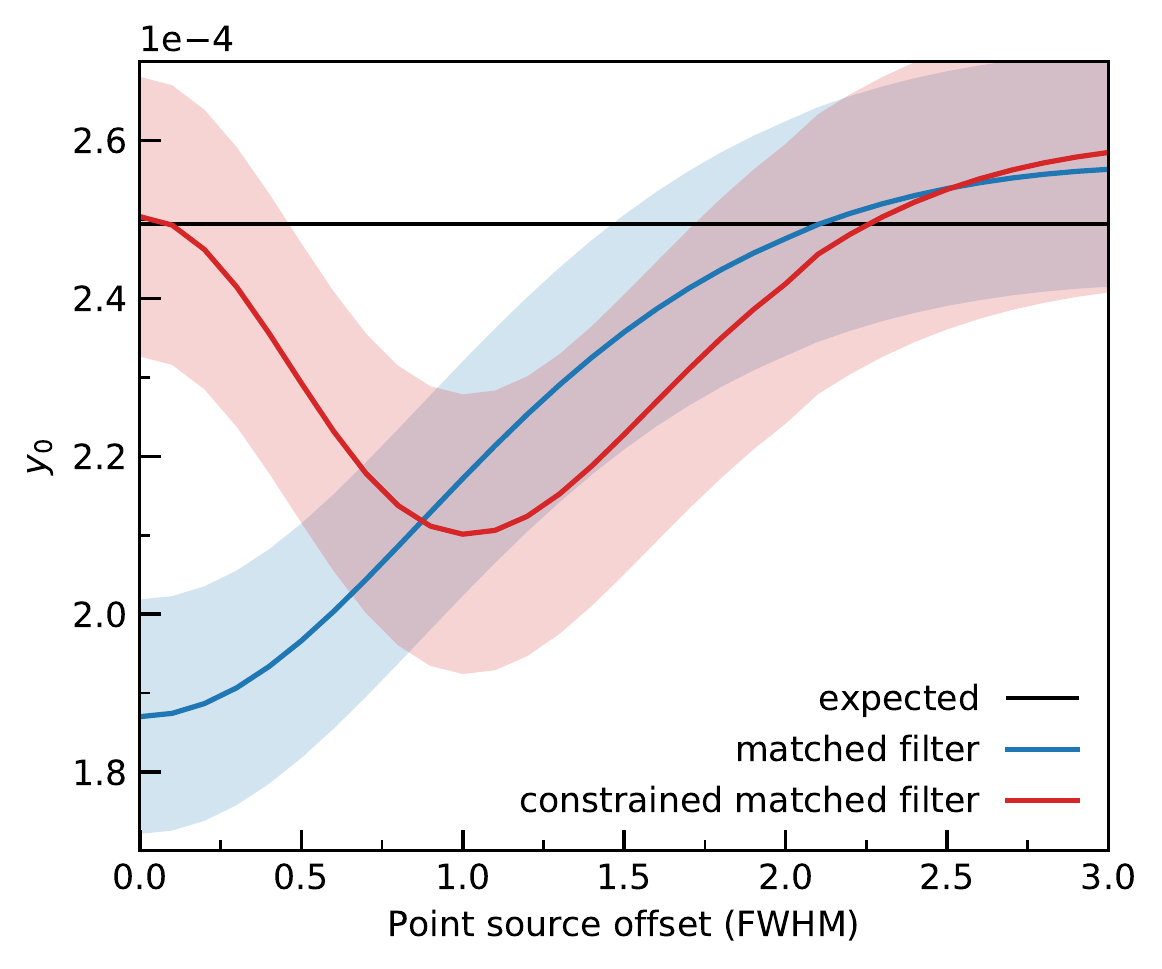}}
  \caption{\textbf{Left-hand panel:} Impact of (radio) point source contamination on the measured Comptonization parameter. We simulate the tSZ decrement at $150 \, \mathrm{GHz}$ for a range of clusters with different masses at a constant redshift of $z=0.2$ using the GNFW profile and add a central point source with a fixed flux density of $10 \, \mathrm{mJy}$ to each of them. The blue data points show the estimates of the central Componization parameter $y_0$ obtained using MFs, while the data shown in red were obtained using CMFs. The solid black line indicates the expected relation. CMFs allows an unbiased measurement of a cluster's flux in the presence of a central point source, while MFs will return a biased value. \textbf{Right-hand panel:} Impact of an offset in the point source location on the previous results. The $x$-axis gives the positional offset relative to the cluster centre. For this test we assume the same resolution, frequency, point source flux, and source redshift, but only consider a single cluster of mass $M_{500} = 10^{15} \, \mathrm{M_\odot}$. The shaded regions indicate the uncertainty of the flux estimates. We find that CMFs perform better than MFs up to an offset of $\sim 0.75 \, \mathrm{FWHM}$. For higher values, both methods provide a similar bias.}
  \label{fig:cmf_results}
\end{figure*}
\begin{figure}
  \centering
  \includegraphics[width=0.48\textwidth]{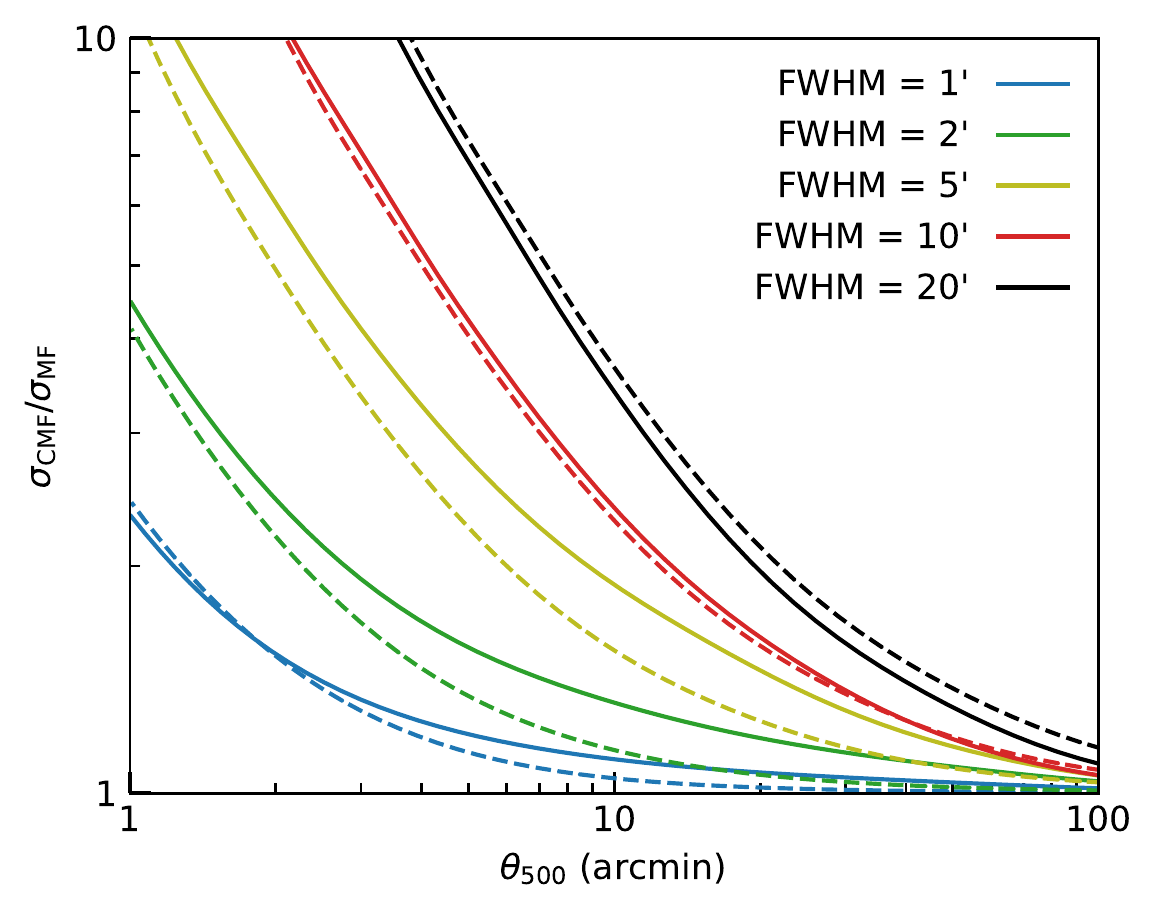}
  \caption{Ratio of the noise level in the CMF and MF filtered maps defined in equation~(\ref{eq:variance}) as a function of cluster size. The different colors correspond to different beam FWHMs. The solid lines correspond to results obtained from the $150 \, \mathrm{GHz}$ mock data while the results shown with dashed lines where obtained using the $350 \, \mathrm{GHz}$ mock data.}
  \label{fig:cmf_noise}
\end{figure}
We restrict our analysis to the extragalactic sky that is relevant for studies of galaxy clusters and cosmological studies by applying a 40 per cent Galactic dust mask to our mock maps. We furthermore exclude the region of the sky that has not been observed by the NVSS in order to keep the properties of our sky model homogeneous. 

All maps are processed at {\small HEALPIX} $n_\mathrm{side}=8192$, which allows to generate mock data with a minimum full width at half-maximum (FWHM) of $1 \, \mathrm{arcmin}$. Maps that come at a lower native resolution are oversampled and smoothed with a narrow Gaussian beam to avoid pixelization artefacts. The microwave sky is simulated at $150 \, \mathrm{GHz}$ and $350 \, \mathrm{GHz}$ with different spatial resolutions ranging from $1 \, \mathrm{arcmin}$ to $20 \, \mathrm{arcmin}$, assuming circular Gaussian beams and white instrumental noise with \hbox{$\sigma^\mathrm{noise}_{150 \, \mathrm{GHz}} = 6.4 \, \mu \mathrm{K_{CMB}}$-arcmin} and \hbox{$\sigma^\mathrm{noise}_{350 \, \mathrm{GHz}} = 25 \, \mu \mathrm{K_{CMB}}$-arcmin}. The wide range of simulated spatial resolutions allows us to test our filtering techniques for instruments ranging from \textit{Planck} to current and future ground-based experiments. The resulting maps are shown in Fig.~\ref{fig:sky_model}.

\subsection{Simulating X-ray data}
\label{sec:xray_data}
In addition to tests using the simulated microwave data that were previously introduced, we apply the single-frequency filters presented in Section \ref{sec:matched_filtering} to simulated X-ray data. We chose to create mock images of the upcoming extended ROentgen Survey with an Imaging Telescope Array (eROSITA, \citealt{Merloni12, Predehl17}), which will explore the high-redshift Universe with unprecedented sensitivity combined with all-sky coverage. This makes eROSITA an ideal case to explore our new filtering techniques, since active galactic nuclei (AGNs) activity is expected to increase with redshift and might dominate the X-ray observed flux around cluster positions (e.g. \citealt{Biffi18, Koulouridis18}).

Following \citet{Clerc18}, the images are simulated in the $0.5$ -- $2 \, \mathrm{keV}$ energy band. Each image has a size of \hbox{$3.6\degr \times 3.6\degr$}, with a pixel size of $4 \, \mathrm{arcsec}$ and a simulated exposure time of $1.6 \, \mathrm{ks}$. We include the X-ray and instrumental background model presented in Table 1 of \citet{Borm14}. A randomly distributed population of point sources, which is described by the \citet{Moretti13} $\log{N}$ -- $\log{S}$ relation, is added. For simplicity, the images only contain a single type of isothermal cluster, simulated using a projected $\beta$--model \citep{Cavaliere76} 
\begin{equation}
	S_\mathrm{X}(\theta) \propto \int_\mathrm{l.o.s} \mathrm{d}l \, n_\mathrm{e}(r)^2 \propto \left[ 1+\left(\frac{\theta}{\theta_\mathrm{c}}\right)\right]^{\frac{1}{2}-3\beta},
\end{equation}
with a fixed flux of $5\times 10^{-13} \, \mathrm{erg \, s^{-1} \, cm{^{-2}}}$, core radius $\theta_\mathrm{c}$ of $20 \, \mathrm{arcsec}$, and $\beta$ of 2/3. All sources are convolved with a Gaussian PSF with an FWHM of $28 \, \mathrm{arcsec}$, which is the expected value in survey mode for eROSITA \citep{Merloni12}. We derived the count rates of the sources from the physical fluxes for a given spectral emission model and using the instrumental response file {\tt erosita\_iv\_7telfov\_ff.rsp}\footnote{The eROSITA response file is available at \\ \hbox{\url{http://www2011.mpe.mpg.de/erosita/response/}}}. In this work, we assume an {\tt APEC} thermal plasma model \citep{Smith01} having a metal abundance of $0.3 \, Z_\odot$ along with a Galactic hydrogen column density corresponding to $1.7\times10^{20} \, \mathrm{cm}^{-2}$ \citep{Kalberla05, Borm14}.
The simulated X-ray images will be used in Section \ref{sec:Xray_application} to demonstrate how CMFs can aid the separation of galaxy clusters and point sources in X-ray surveys. 

\section{Results}
\label{sec:results}

\subsection{Photometry of clusters with a central point source}
\label{sec:photometry}

Using the simulation pipeline introduced in the previous section, we first investigate how the new filtering technique presented in this work can improve the photometry of clusters that harbour a bright central point source. We do so by creating $150 \, \mathrm{GHz}$ mock observations of clusters with masses ranging from $10^{14} \, \mathrm{M}_\odot$ to $10^{15} \, \mathrm{M}_\odot$ at a constant redshift of $0.2$. Each simulated cluster features a central radio source with a fixed flux density of $10 \, \mathrm{mJy}$ at $150 \, \mathrm{GHz}$. The beam is assumed to have an FWHM of $1 \, \mathrm{arcmin}$. The values of the central Comptonization parameter computed from the measured cluster flux after filtering are shown in the left-hand panel of Fig.~\ref{fig:cmf_results}. By construction, a CMF always returns an unbiased result, while the values obtained through MF are biased low with a linear dependence on the brightness of the point source. For the given flux density this bias has a significance of $\sim 4  \sigma$ and increases to $\sim 5  \sigma$ for lower cluster masses due to the decreasing cluster size. 
The biased fluxes can therefore lead to a non-detection of low-mass clusters and biased inferred cluster properties for high-mass systems.

We also consider a potential offset of a bright central point source relative to the cluster centre. If a central point source is not aligned with the cluster, both methods will find a bias due to ringing artefacts around the filtered point source. We find that for small angular separations up to $\sim 0.75 \, \mathrm{FWHM}$ CMFs return a value with a bias that is smaller than the one observed in the values returned by MFs, while both methods find similar values for larger offsets.

The comparison above highlights a clear advantage of the CMF, which however is bought with an increase in the noise level in the filtered map that limits the usefulness of the method in some cases. This noise increase results from placing additional constraints that inevitably lower the degrees of freedom available for the optimization of the variance of the filtered map. Figure~\ref{fig:cmf_noise} shows the ratio of the noise in the filtered maps as a function of apparent cluster size and instrument beam. This ratio scales linearly with the cluster-size-to-beam ratio if the map noise is Gaussian. The differences between the results obtained at $150 \, \mathrm{GHz}$ and $350 \, \mathrm{GHz}$ are due to the different foreground properties. We find that the CMF provides maps with a marginally increased noise level for most modern ground-based mm telescopes that offer a typical resolution of $\sim 1 \, \mathrm{arcmin}$. The use case for low-resolution instruments like \textit{Planck} is however restricted to large, mostly nearby clusters with radii of several tens of arcminutes.

\subsection{Application to \textit{Planck} data}
In addition to tests on simulated microwave images we apply all filters presented in Section~\ref{sec:matched_filtering} to \textit{Planck} HFI data of the Perseus galaxy cluster at $z = 0.0179$. The brightest cluster galaxy (BCG) of the Perseus cluster (NGC 1275) is a powerful radio source known as Perseus A that is unresolved in all \textit{Planck} bands. While the MMF and CMMF are applied directly to the HFI data without any pre-processing other than converting the $545$ and $857 \, \mathrm{GHz}$ maps to units of $\mathrm{K_{CMB}}$, the HFI maps are combined into a single map before applying the single-frequency filers. This is achieved by smoothing the maps to a common resolution of $9.68 \, \mathrm{arcmin}$ after which they are combined into a $y$-map with ILC or constrained ILC (CILC) algorithms (\citealt{Remazeilles11a}; see Appendix C of \citealt{Erler18} for details). The radio galaxy Perseus A appears as a bright source with negative amplitude in the ILC $y$-map due to its diminishing brightness with increasing frequency, which is also seen in the MILCA and NILC $y$-maps published by the \citet{Planck_y}. In contrast, the CILC algorithm allows to constrain the Perseus A SED and thus remove its contamination to the $y$-map. The Perseus A SED used for the CILC and CMMF algorithms is extracted directly from the \textit{Planck} HFI data using CMFs that remove the tSZ contamination by the ICM of the cluster and found to be well approximated by a power-law with spectral index $\alpha = 0.78 \pm 0.05$ (see Table~\ref{tab:SED}). We model the tSZ signal of the Perseus cluster with a GNFW pressure profile with $\theta_{500} = 59.7 \, \mathrm{arcmin}$ \citep{Urban14} and use a non-relativistic approximation of the tSZ spectrum. All maps are \hbox{$10\degr \times 10\degr$} fields centred on $(\mathrm{RA, \, Dec.})=(03\mathrm{h}19\mathrm{m}47.2\mathrm{s}, \, +41\degr 30\arcmin 47\arcsec)$.
\begin{table}
\begin{center}
\begin{tabular}{ccc}
\hline
$\nu$ & FWHM & $S_\nu$ \\
(GHz) & (arcmin) & (Jy) \\ \hline
100 & 9.68 & 10.36 $\pm$ 0.15 \\
143 & 7.30 & \,~7.80 $\pm$ 0.13 \\
217 & 5.02 & \,~5.74 $\pm$ 0.25 \\
353 & 4.94 & \,~4.12 $\pm$ 0.87 \\
545 & 4.83 & \,~2.82 $\pm$ 2.81 \\
857 & 4.64 & \,~2.12 $\pm$ 7.90 \\ \hline
\end{tabular}
\end{center}
\caption{SED of Perseus A extracted from \textit{Planck} HFI data using CMFs that remove the tSZ signal of the cluster. It is well approximated by a power-law with a spectral index of $0.78 \pm 0.05$. In turn this SED is used to clean tSZ maps of the Perseus cluster in various ILC and MMF approaches.}
\label{tab:SED}
\end{table}

\begin{figure*}
  \centering
  \includegraphics[width=0.9\textwidth]{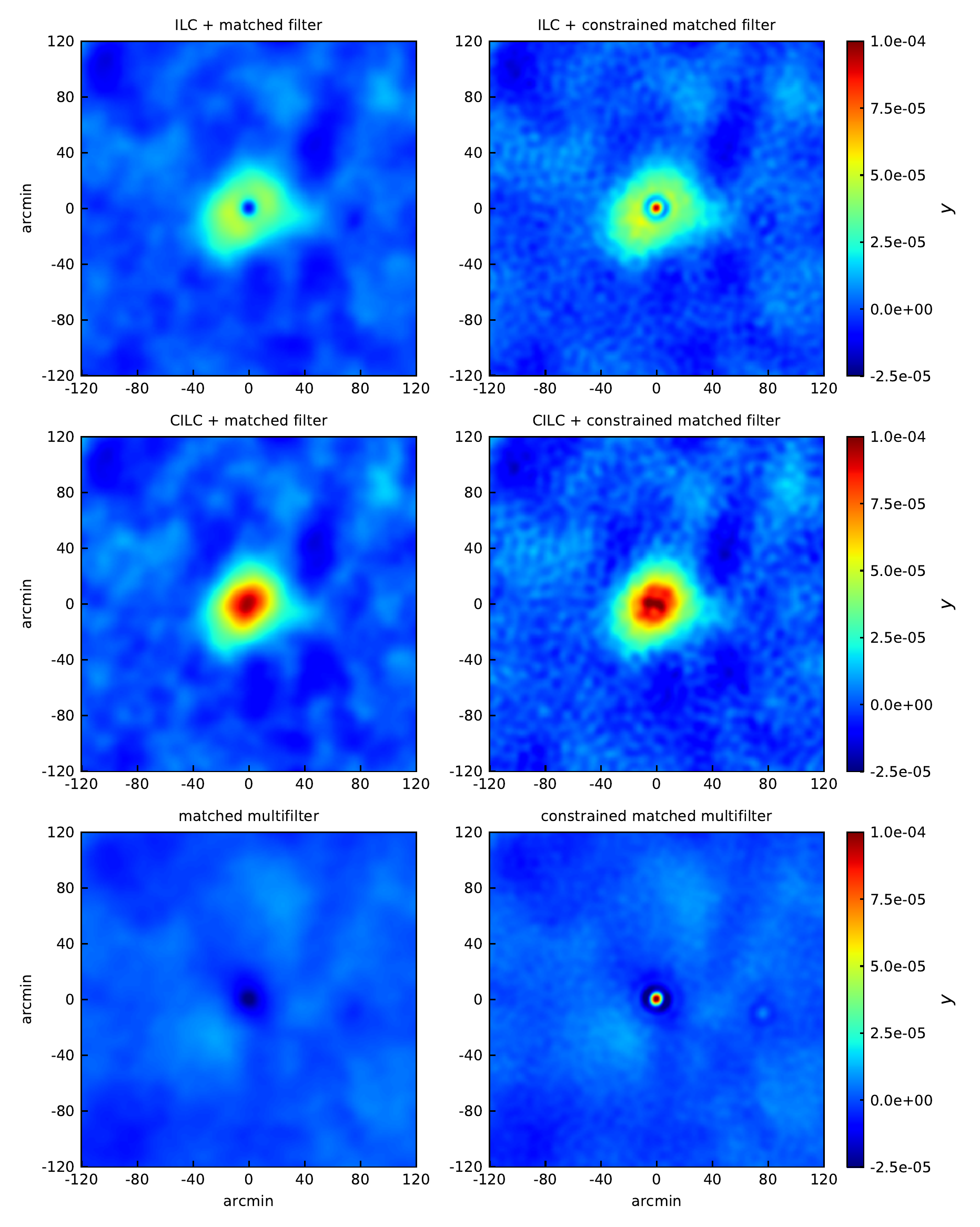}
  \caption{Filtered maps of the Perseus galaxy cluster processed with the filtering techniques presented in Section~\ref{sec:matched_filtering}. The BCG of Perseus hosts a bright radio source called Perseus A that is known to contaminate tSZ observations of the cluster, leading to biased fluxes. We applied all algorithms to $10\degr \times 10\degr$ \textit{Planck} HFI maps centred on $(\mathrm{RA, \, Dec.}) = (03\mathrm{h}19\mathrm{m}47.2\mathrm{s}, \, +41\degr 30\arcmin 47\arcsec)$. The maps above show the inner $4\degr \times 4\degr$ of the field. In order to apply the single-frequency filters, the six HFI maps were combined into a $y$-map using ILC and CILC algorithms, the latter of which allows to remove the contamination caused by Perseus A by constraining its SED. Our comparison shows that there are various ways of removing point sources from clusters using either spectral or spatial constraints or a combination of both, all of which find consistent values for the central Comptonization parameter $y_0$ and thus $Y_{500}$. The best SNR is delivered by the CMMF, which yields a value of 24. This is comparable to the SNR of clusters with similar mass and redshift to Perseus that do not suffer from point source contamination, like the Coma cluster.}
  \label{fig:Perseus}
\end{figure*}

We summarize our results by providing the extracted values for the central Comptonization parameter $y_0$ and the derived integrated value $Y_{500}$ in Table~\ref{tab:Perseus_results}. The latter is integrated in a cylindrical aperture with the radius $\theta_{500}$
\begin{equation}
	Y_{500}^\mathrm{cyl} = y_0 \frac{2 \pi}{(10^6 \, pc)^2}\int_0^{D_\mathrm{A}\theta_{500}} \mathrm{d}r \, y(r) \, r,
\end{equation}
where $y(r)$ is the cluster template that has been normalized to unit amplitude and $D_\mathrm{A}$ is the angular diameter distance of the cluster. The processed maps are shown in Fig.~\ref{fig:Perseus}. 

If neither a spatial nor a spectral constraint for Perseus A is used, as in the MMF and ILC + MF scenarios, we extract a strongly biased negative value for $y_0$ and thus $Y_{500}$. This provides a plausible explanation for the necessity of point source masks that are the reason why the Perseus cluster is not listed in the \textit{Planck} SZ cluster catalogues (PSZ and PSZ2, \citealt{Planck13, Planck16}), which were built using two MMF pipelines ({\small MMF1} and {\small MMF3}) and the Bayesian {\small PowellSnakes} ({\small PwS}) algorithm. 

The bias introduced by Perseus A is removed by applying a CMF to the same $y$-map, which yields $y_0 = (9.4 \pm 0.7) \times 10^{-5}$. For the application of a CMF to an ILC $y$-map it is critical to smooth all maps to a common resolution before combining them. Combining the maps in Fourier space at their native resolution will distort the beam in the $y$-map, which increases the complexity of constraining the spatial template of the beam for point source removal.

Using the Perseus A SED to construct a CILC $y$-map before filtering is an alternative way to remove the bias introduced by the radio source. In that case, both the MF and the CMF yield similar values for $y_0$, both of which are consistent with the previous result. Placing a spectral constraint in the CILC step however results in a noisier $y$-map and thus a slightly lower SNR in both cases.

Finally, applying a CMMF that uses both the SED of Perseus A and our knowledge of the \textit{Planck} beams yields $y_0 = (10.0 \pm 0.42) \times 10^{-5}$, which is in agreement with the previous values and with an SNR of 24 offers the clearest signal of all methods compared here. This SNR is comparable to the SNR of 22 we obtain by applying an MMF to \textit{Planck} HFI maps of the Coma cluster, a system of similar mass at z= 0.0231. Using the $M_{500}-Y_{500}$ scaling relation from the \citet{Planck13_cosmo, Planck16_cosmo} and converting to $Y_{500}^\mathrm{sph}$ we find a mass of $(6.97 \pm 0.24)\times 10^{14} \, \mathrm{M}_\odot$ for the Perseus cluster, which is consistent with the value obtained by \citet{Urban14}\footnote{The error on the mass includes the uncertainties of the scaling relation parameters given by \citet{Planck16_cosmo}, which we assume to be uncorrelated.}.

\begin{table}
\begin{center}
\setlength{\tabcolsep}{5pt}
\begin{tabular}{cccc}
\hline
Technique & $y_0$ & $Y_{500}$ & SNR \\
 & $10^{-5}$ &  $10^{-5} \, \mathrm{Mpc}^2$ &  \\ \hline
~~ILC + MF   & \,~-0.74 $\pm$  0.66 & -0.50 $\pm$  0.44  & -1.1 \\
~~~~ILC + CMF  & \,~~9.35  $\pm$  0.70 & ~6.31  $\pm$  0.47  & 13.4 \\
CILC + MF  & \,~~9.44  $\pm$  0.77 & ~6.37  $\pm$  0.52  & 12.3 \\
~~CILC + CMF & \,~~9.77  $\pm$  0.82 & ~6.59  $\pm$  0.55  & 12.0 \\
~~~MMF        & \,~-2.64 $\pm$  0.39 & -1.77 $\pm$  0.27 & -6.8 \\
~~~CMMF       & \,~~10.0 $\pm$  0.42 & ~6.76 $\pm$  0.28  & 24.0 \\ \hline
\end{tabular}
\end{center}
\caption{Comparison of the extracted tSZ signal of the Perseus galaxy cluster extracted from \textit{Planck} HFI data with various ILC and MMF techniques. The corresponding maps are shown in Fig.~\ref{fig:Perseus}. The ILC-based techniques first combine the six HFI maps in an optimal linear combination, after which we apply either a MF or a CMF. The MMF techniques are directly applied to the HFI maps. The CILC and CMMF techniques use the Perseus A SED given in Table~\ref{tab:SED}. Radio sources like Perseus A will appear as sources with negative $y$ in ILC $y$-maps and MMF maps, leading to biased photometry if not accounted for.}
\label{tab:Perseus_results}
\end{table}

For Coma, all six methods yield similar values for $y_0$ due to the lack of a bright central radio or FIR source. We find however that the two multifilters deliver an almost identical SNR as the ILC plus MF techniques, while the CILC approach gives a slightly lower SNR of 17. This indicates that the additional constraints are `cheaper' for multifilters but come at the drawback that multiple constraints have to be placed for sources with identical spatial template but different SEDs. Combining an ILC map and CMFs will remove sources just based on their spatial signature with no need to have constraints on their SED. 

This example illustrates that there are multiple ways of dealing with point source contamination in clusters. The advantage of the CMF over using spectral constraints is that it is often easier to characterize the instrument beam than measuring the SED of a source. Radio sources like Perseus A can show variability and extrapolating their fluxes to microwave frequencies based on radio measurements often relies on the assumption of a perfect power-law SED, which can be prone to mistakes since many sources are known to have SEDs that deviate from a power-law \citep{Herbig92}. Furthermore, using spectral information will require individual measurements for each source, while a spatial technique can be applied blindly to a large number of objects. 

\subsection{Blind cluster detection and X-ray application}
\label{sec:Xray_application}
\begin{figure*}
  \centering
  \includegraphics[width=1.0\textwidth]{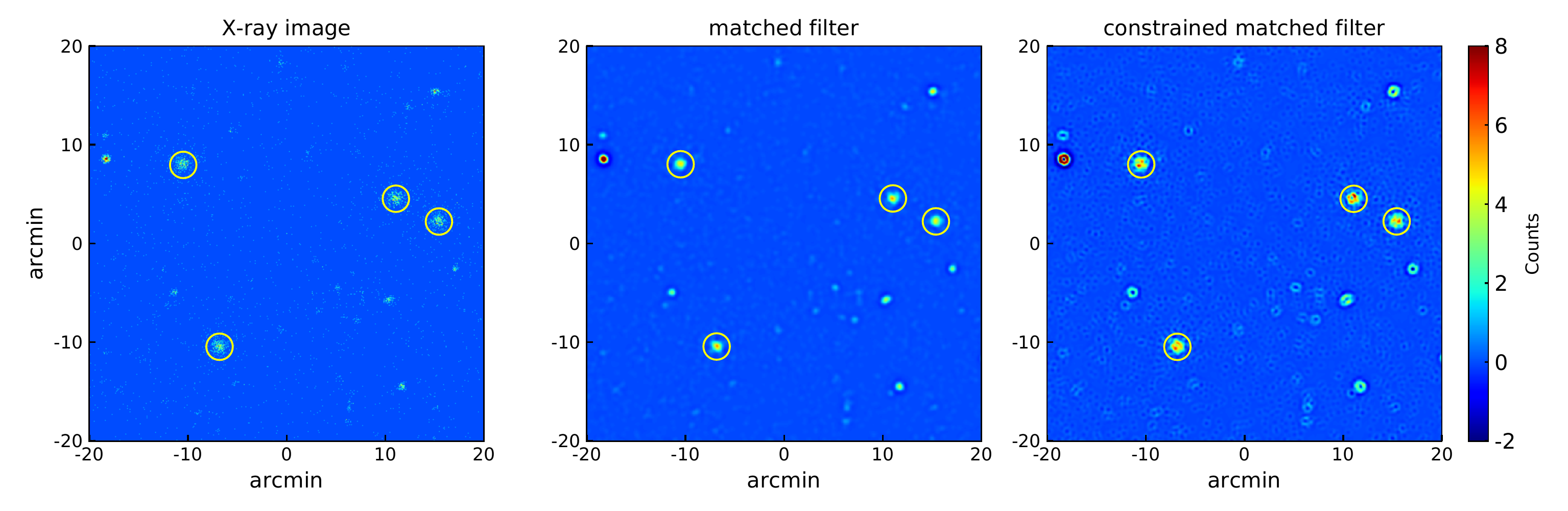}
  \caption{Zoom-in on a simulated X-ray photon image (left) and the same image convolved with an MF (centre) and CMF (right). The colour bar to the right has been cropped at eight counts to highlight faint structures in the filtered maps. The mock data features realistic X-ray and instrumental backgrounds as well as a realistic point source population, but for simplicity only contains multiple realizations of a single simulated cluster (highlighted with yellow circles). A detailed description of the data can be found in Section~\ref{sec:xray_data}. The MF yields an SNR amplification of the clusters but its response to point sources is similar to that of clusters, which makes the separation of the two source populations challenging in some cases. In contrast, the CMF nullifies point sources and leaves behind `doughnuts' at their positions in the map. Since both filters have an identical response to clusters, combining their results allows for a quick and simple separation of the two source populations, which is shown in Fig.~\ref{fig:X-ray}.}
  \label{fig:xray_image}
\end{figure*}
We also investigate the potential application of the CMF to reduce point source contamination for blind cluster detection. In tSZ surveys below $217 \, \mathrm{GHz}$ point sources will not be misclassified as galaxy clusters due to the tSZ effect's characteristic decrement. They can however lower the decrement or even overpower it, which can lead to a biased flux or a non-detection as has been illustrated previously in Section~\ref{sec:photometry}. At higher frequencies, in the tSZ increment, point sources can bias the flux and might be misclassified as clusters. For instruments like \textit{Planck} the situation has been mitigated by multifrequency coverage (e.g. \citealt{Bartlett06}), but prominent examples like the Perseus cluster remain. 

Point source contamination is an even greater issue in X-ray surveys due to the stochastic nature of the observed signal. The upcoming eROSITA survey is expected to detect about $100\,000$ galaxy clusters \citep{Pillepich12, Clerc18} as well as millions of AGNs. Separating both source populations presents a major challenge for cluster detection algorithms. The CMF introduced here presents an additional tool for this task that has the benefit of using reasonable assumptions, like well-known cluster profiles and the PSF of the instrument, to deliver an optimal result. In the remaining part of this section we will provide a brief outline how the traditional and CMFs can be combined to detect clusters in X-ray surveys and reduce the number of misclassified point sources.  

\begin{figure}
  \centering
  \includegraphics[width=0.48\textwidth]{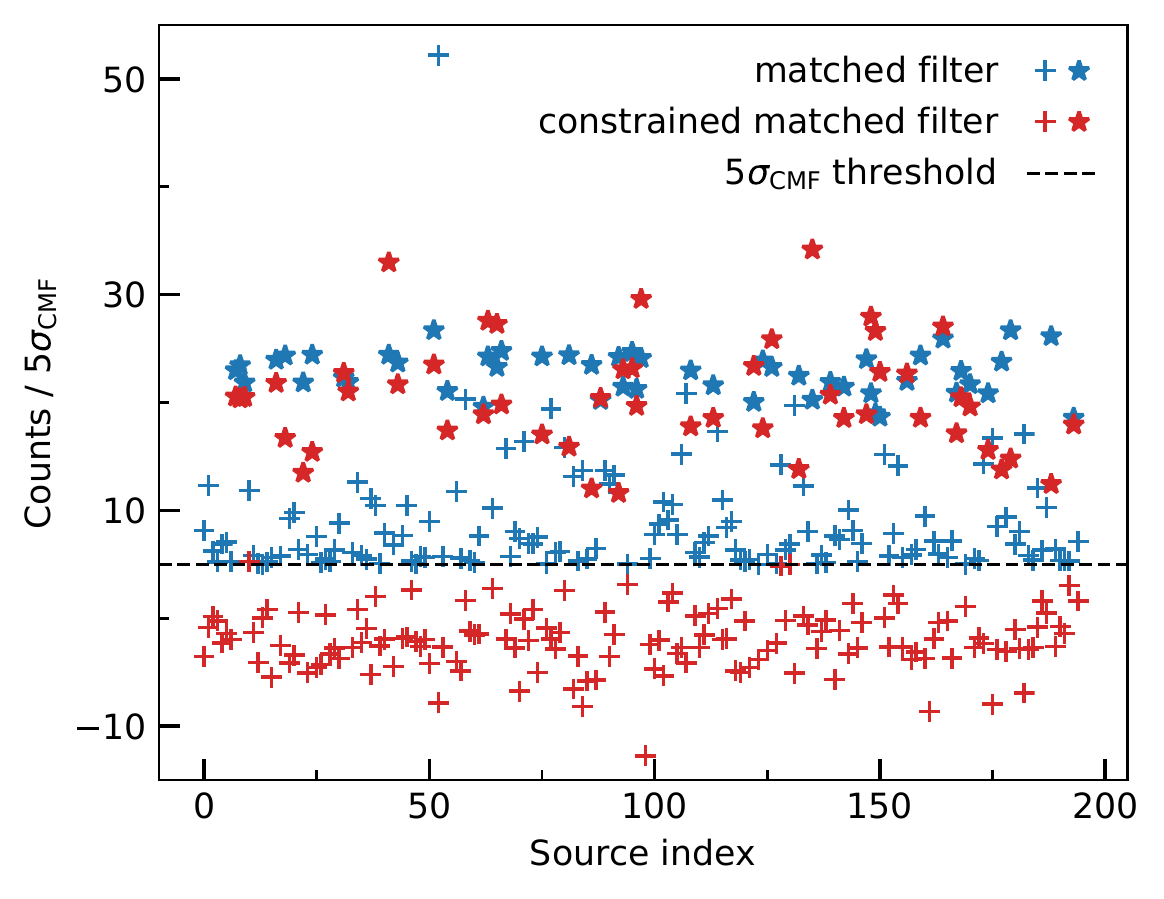}
  \caption{A simplified demonstration of the application of the CMF to eROSITA mock data.  Simulated clusters are shown as stars while point sources are shown as crosses. Applying an MF to a map with $1.6 \, \mathrm{ks}$ exposure time will typically result in 200 sources with centroid values above $5\sigma_\mathrm{CMF}$. When measuring the corresponding values in a map processed with a CMF, the vast majority of previously detected point sources (blue crosses) will drop below the threshold (red crosses), leaving us with a cleaner sample of cluster candidates.}
  \label{fig:X-ray}
\end{figure}

We perform our tests on the eROSITA mock data that was introduced in Section~\ref{sec:xray_data}. Each field is  filtered with both an MF and a CMF. We then apply a simple source finder\footnote{We use the {\tt find\_peaks()} function of the python {\small PHOTUTILS} package.} to the former map to identify bright sources above a fixed threshold, e.g. $5\sigma_\mathrm{CMF}$ as defined by equation~(\ref{eq:variance}), and determine their centroids. This typically leaves us with around 200 source candidates, the majority of which are point sources. We then determine the values of the map processed with the CMF at the position of the previously measured centroids and only classify objects for which both values lie above the former threshold as cluster candidates. This procedure is illustrated for a single field in Figs.~\ref{fig:xray_image} and~\ref{fig:X-ray} . 

We find that using both filters in conjunction will strongly reduce the number of misclassified point sources. As demonstrated clearly in Fig.~\ref{fig:X-ray}, the CMF yields a better segregation of point sources in terms of their SNR but also raises the scatter of the filtered cluster photon counts due to the increased map noise.

\section{Discussion}
\label{sec:discussion}

The new CMF and CMMF techniques presented in this work are straightforward extensions of the MF concept that enable optimal extraction of sources with known templates while  at the same time allowing for an optimal reduction of known contaminating sources. The results presented in Section~\ref{sec:results} focused on the reduction of point source contamination to SZ and X-ray observations of galaxy clusters, but it is important to stress that the methods presented here are applicable to any contaminating source that can be approximated through a known template. It is also possible to place more than one constraint, yet care has to be taken since every additional constraint will result in a noisier map. As with any MF, the values found in the filtered map will be biased if the source template does not match the true shape of a resolved source. We note however that the CMFs can provide a slightly larger bias than the traditional MF if the desired source is more compact than its template and other compact sources are supposed to be removed.  

A technique similar to the CMMF presented in this work was explored by \citet{Herranz05}, who derived an unbiased MMF to minimize the contamination of the tSZ to kSZ maps and vice versa. These authors derived a two-component version of the filter presented here and then use the same spatial but different spectral templates for the two different SZ components to separate them. However, a potential drawback of this method is that the spatial templates of the tSZ and kSZ signals should in general be different, especially for merging systems.

An important detail of the new methods is their dependence on the spatial resolution of the instrument, which has a crucial impact on the noise level of the filtered map. Compact clusters will thus remain spatially indistinguishable from point sources if the instrument beam is large. This also restricts the application of the constrained filters on \textit{Planck} data to nearby clusters with large apparent radii. The situation improves when the instrument beam has an FWHM of $\sim 1 \, \mathrm{arcmin}$ or less, at which point the noise will only increase by a few percent compared to a matched filtered map for most cluster sizes. Such resolution is quite common for ground-based cluster surveys like the ones performed by the SPT and ACT. However, additional filtering will be applied for ground-based instruments to reduce atmospheric contamination. The impact of these filtering steps on the astrophysical signal has to be understood and characterized before MFs are applied (e.g. \citealt{Bleem15}).

The new filtering techniques are especially interesting for studies of the kSZ and relativistic tSZ at sub-mm wavelengths with upcoming instruments like CCAT-prime\footnote{\url{http://www.ccatobservatory.org/}}. CCAT-prime will be a $6 \, \mathrm{m}$ diameter submillimetre survey telescope that is going to operate at $5600 \, \mathrm{m}$ altitude on the summit of the Cerro Chajnantor in the Chilean Atacama Desert \citep{Parshley18a, Parshley18b}. The high and dry site offers superb conditions for observations at frequencies ranging from $270 \, \mathrm{GHz}$ to $860 \, \mathrm{GHz}$ \citep{Vavagiakis18} at up to one order of magnitude better sensitivity than \textit{Planck} \citep{, Erler18,Mittal18}. Combined with mm-data of the advanced ACT-pol survey, CCAT-prime will offer full coverage of the SZ spectrum and allow significant improvements over \textit{Planck} in measuring cluster parameters  \citep{Erler18, Stacey18}. In order to constrain key properties of clusters via the SZ effects, accurate mm and sub-mm photometry will be required and MF techniques including the ones introduced here are an excellent tool for this \citep{Soergel17, Erler18}.

One of the most important applications of the CMF will be next-generation wide-area X-ray surveys, such as eROSITA, that aim to detect the diffuse emission of many thousands of galaxy clusters out to high redshift in the presence of millions of AGNs. The need for new techniques for better point source separation was recently highlighted by \citet{Biffi18}, who used X-ray mocks derived from the hydrodynamical \textit{Magneticum Pathfinder Simulation} to investigate the contribution of AGNs inside clusters to the X-ray luminosity of the ICM. The methods presented in this work are especially tailored to this application, since they only require a spatial template and provide an optimal and unbiased result. An important benefit of the filters presented here is their ability to separate clusters and point sources even if they are aligned. On one hand, this can lead to biased photometry of clusters with compact cool cores if the template does not account for it, but on the other hand such a bias can actually be useful to mitigate the so-called cool-core selection bias in X-ray cluster surveys.

We note however that the CMF should not be considered as a replacement for well-proven and tested methods but rather presents an additional tool that will work best in conjunction with other methods such as the traditional MF or e.g. the well-known sliding cell \citep{Harnden84} and wavelet \citep{Freeman02} algorithms, since significant discrepancies between their extracted signals hint at potential point source contamination.

The X-ray analysis presented here was deliberately chosen to be qualitative and focuses on the conceptual application of the new methods, since we do not account for the Poissonian statistics that govern X-ray observations and do not tune the detection threshold to maximize the number of detected clusters while staying below a fixed rate of spurious detections. In addition to a robust X-ray implementation of the filters, future, more quantitative studies of the X-ray application of the CMF should include a realistic energy and line-of-sight dependent PSF and tests using archival X-ray data.

Other recent attempts on improving the separation of point sources and galaxy clusters in X-ray data sets include the combination with optical data \citep{Green17} and a new MMF technique introduced by \citet{Tarrio16, Tarrio18} who used ROSAT data as an additional \textit{Planck} channel to use the different point source populations in the two data sets.

\section{Conclusions}
\label{sec:conclusion}

This work introduced a new way to generalize MFs and MMFs to separate desired and undesired sources based on just their spatial (CMF) or their spatial and spectral (CMMF) characteristics. Adding additional constraints will reduce the SNR of the sources, but if both source and contaminant are well approximated by given templates the methods introduced here will allow for unbiased photometry and reduced confusion. When applied to Gaussian data, MFs are optimal in the least-square sense, making them ideal tools for the extraction of the SZ signal of galaxy clusters from microwave data. However, traditional MF techniques can perform poorly if microwave data of galaxy clusters are contaminated by point sources.

At microwave frequencies, there are two distinct populations of point-like sources that are spatially correlated with galaxy clusters. The first consists of radio-bright AGNs that are found at the centres of many BCGs, and the second being composed of dusty star-forming galaxies.
Using realistic microwave mock data we showed that the CMF introduced in this work allows for unbiased photometry of clusters that harbour a central point source. If applied at multiple frequencies it enables studies of the SZ spectrum of clusters with no need to account for the SED of the point source. We showed that our method requires sufficient spatial resolution to be competitive and otherwise will yield an unbiased but noisy result.
Applying constrained and unconstrained MFs and MMFs to \textit{Planck} HFI data of the Perseus cluster, which features a bright central radio source, demonstrated that there are multiple ways to remove a central source from actual data, requiring only spatial or spectral constraints, or the combination of both. In the latter case we showed that Perseus can be detected with an SNR typical for a cluster of its mass and redshift. However using only spatial constraints will reduce contamination by point sources regardless of their SED. 

The application of the methods presented here is especially interesting to the upcoming CCAT-prime and eROSITA cluster surveys. While CCAT-prime will benefit from unbiased photometry of clusters with central point sources for detailed measurements of the rSZ and kSZ effects, point source confusion during cluster detection is a major concern for X-ray surveys. We illustrated how the CMF can provide an optimal way to distinguish between clusters and point sources and showed that the new method has the potential to be developed into a competitive cluster finding algorithm.

\section*{Acknowledgements}

The authors would like to thank the anonymous referee for their valuable feedback, as well as Jean-Baptiste Melin, Paula Tarr{\'{\i}}o, Thomas Reiprich, Florian Pacaud, Nicolas Clerc, Eve Vavagiakis, Christos Karoumpis and Sandra Unruh for insightful comments and discussions.
JE, KB and FB acknowledge partial funding from the Transregio programme TRR33 of the Deutsche Forschungsgemeinschaft (DFG). JE furthermore acknowledges  support  by  the  Bonn-Cologne  Graduate  School  of  Physics and  Astronomy (BCGS)
MERC acknowledges support by the German Aerospace Agency (DLR) with funds from the Ministry of Economy and Technology (BMWi) through grant 50 OR 1514.
The simulations of the CIB used in this paper were developed by the WebSky Extragalactic CMB Mocks team, with the continuous support of the Canadian Institute for Theoretical Astrophysics (CITA), the Canadian Institute for Advanced Research (CIFAR), and the Natural Sciences and Engineering Council of Canada (NSERC), and were generated on the GPC supercomputer at the SciNet HPC Consortium. SciNet is funded by the Canada Foundation for Innovation under the auspices of Compute Canada, the Government of Ontario, Ontario Research Fund -- Research Excellence, and the University of Toronto. This research used {\small PHOTUTILS} and {\small ASTROPY}, a community-developed core {\small PYTHON} package for Astronomy \citep{Astropy18}. 
%
%







\appendix

\section{all-sky formalism}
\label{sec:allsky}

The MF formalism presented in Section~\ref{sec:matched_filtering} used the flat sky approximation but can be adopted to the full sphere with little effort. Implementing MFs on the full sphere can have advantages in certain situations, because we can avoid using an approximate projection to a flat-sky geometry. \citet{Schaefer06} provide an excellent overview on the details. This section is intended to give a summary of the most important points.

Assuming radial symmetry of the sources that we are interested in (i.e. $m=0$) and using the convolution theorem on the sphere we can relate the spherical harmonic coefficients of the unfiltered map $a_\mathrm{\ell m}^\mathrm{unfilt}$ to the ones of the filtered map $a_\mathrm{\ell m}^\mathrm{filt}$ by:
\begin{equation}
   a_\mathrm{\ell m}^\mathrm{filt} = \sqrt{\frac{4 \pi}{2\ell + 1}} \, \Psi_{\ell 0} \, a_\mathrm{\ell m}^\mathrm{unfilt} \equiv F_\ell \, a_\mathrm{\ell m}^\mathrm{unfilt}.
   \label{eq:conv_theorem}
\end{equation}
The new all-sky MF $\bm{F}$ will thus be 
\begin{equation}
  \bm{F} = (\bm{\tau}^\mathrm{T} \mathbfss{C}^{-1} \bm{\tau})^{-1}\tilde{\bm{\tau}}\mathbfss{C}^{-1},
\end{equation}
where $\mathbfss{C}$ is the power spectrum of the all-sky map recast as a diagonal matrix as was done in Section~\ref{sec:matched_filtering} and the elements of $\bm{\tau}$ and $\tilde{\bm{\tau}}$ are defined as:
\begin{equation}
   \tau_\ell = \sqrt{\frac{2\ell+1}{4 \pi}} \cdot  {\tilde \tau}_{\ell 0} = \sqrt{\frac{2\ell+1}{4 \pi}} \cdot y_{\ell0} \cdot B_{\ell} \cdot w_{\ell}.
   \label{eq:tau_l}
\end{equation}
Here, $y_{\ell0}$ denotes the spherical harmonic transform of the source template profile, while $B_{\ell}$ and $w_{\ell}$ are the beam and pixel window functions. When computing the $C_\ell$ it is often useful to mask the brightest regions of the Galaxy to reduce contamination from bright ringing artefacts and ensure that the data are Gaussian.

The CMF can be applied to the full sphere analogously. Using equation~(\ref{eq:conv_theorem}) the all-sky filter can be written as:
\begin{equation}
  \bm{F} = \bm{e}^\mathrm{T}\left(\mathbfss{T}^\mathrm{T} \mathbfss{C}^{-1}\mathbfss{T} \right)^{-1} \tilde{\mathbfss{T}} \mathbfss{C}^{-1}.
\end{equation}
As defined in Section~\ref{sec:matched_filtering}$, \mathbfss{T}$ and $\tilde{\mathbfss{T}}$ are matrices build from the $n$ spatial constraints
\begin{equation}
  \mathbfss{T}=\begin{pmatrix}
    \tau_1[1]  & \tau_2[1] & \dots & \tau_n[1]  \\
    \vdots      & \vdots       & \ddots& \vdots \\
    \tau_1[n_\ell] & \tau_2[n_\ell] & \dots & \tau_n[n_\ell] \\
    \end{pmatrix},
\end{equation}
\begin{equation}
  \tilde{\mathbfss{T}}=\begin{pmatrix}
    \tilde{\tau}_1[1]  & \tilde{\tau}_2[1] & \dots & \tilde{\tau}_n[1]  \\
    \vdots      & \vdots       & \ddots& \vdots \\
    \tilde{\tau}_1[n_\ell] & \tilde{\tau}_2[n_\ell] & \dots & \tilde{\tau}_n[n_\ell] \\
    \end{pmatrix},
\end{equation}
where the components $\tau_i$ and $\tilde{\tau}_i$ are defined for each template $i$ as done in equation~(\ref{eq:tau_l}).



\bsp	
\label{lastpage}
\end{document}